\documentclass[aip,pof, reprint,floatfix,onecolumn]{revtex4-1}
\usepackage{graphicx}
\usepackage{bm}
\usepackage{natbib}
\usepackage{subfigure}
\usepackage{epstopdf}
\usepackage{amssymb,amsmath}
\usepackage{color} 
\usepackage[abs]{overpic}
\usepackage[overlay,absolute]{textpos}
\usepackage{color}

\textblockorigin{-5mm}{0mm}   

\begin{document}
\title{Similarity and singularity in adhesive elastohydrodynamic touchdown}
\author{Andreas Carlson}
\email{carlson@seas.harvard.edu}
\affiliation{Paulson School of Engineering and Applied Sciences and Wyss Institute for Biologically Inspired Engineering, Harvard University, Cambridge, USA.}

\author{L. Mahadevan$^{1,}$}
\email{lm@seas.harvard.edu}
\affiliation{Kavli Institute for Bionano Science and Technology, Departments of Physics, and Organismic and Evolutionary Biology, Harvard University, Cambridge, USA.}

\date{\today}

\begin{abstract}
We consider the touchdown of an elastic sheet as it adheres to a wall, which has a dynamics that is limited by the viscous resistance provided by the squeeze flow of the intervening liquid trapped between the two solid surfaces. The dynamics of the sheet is described mathematically by elastohydrodynamic lubrication theory, coupling the elastic deformation of the sheet, the microscopic van der Waals adhesion and the viscous thin film flow. We use a combination of numerical simulations of the governing partial differential equation and a scaling analysis to describe the self-similar solution of the touchdown of the sheet as it approaches the wall. An analysis of the equation satisfied by the similarity variables in the vicinity of the touchdown event shows that an entire sequence of solutions are allowed. However, a comparison of these shows that only the fundamental similarity solution is observed in the time-dependent numerical simulations, consistent with the fact that it alone is stable. Our analysis generalizes similar approaches for rupture in capillary thin film hydrodynamics and suggests experimentally verifiable predictions for a new class of singular flows linking elasticity, hydrodynamics and adhesion.
\end{abstract}

\maketitle
The mathematical description of fluid flow \cite{Batchelor} often leads to diverging velocity gradients in bulk flows or geometric singularities in free surface flows \cite{Barenblatt,eggers:1999,eggers:2009}, with implications for many applications such as microfluidics, coating, food processing etc. Here, we describe the elastohydrodynamic thin film flow associated with van der Waals-driven touch-down between an elastic sheet and a solid substrate, which is inspired by the rupture of capillary thin fluid films on substrates \cite{jain:1976,william:1982,Bruelbacht:1988, Bernoff:1998,zhang:1999,Witelski:1999, Witelski:2000,Vaynblat:2001,Witelski:2010,craster:2009} that reveal a finite--time singularity\cite{zhang:1999,Witelski:1999}.


Our analysis focuses on small scale flows wherein inertial effects from both the fluid and the sheet may be neglected. Furthermore, we consider the touchdown of a long and wide elastic sheet with a length $L$, which is initially separated from a solid substrate by a thin viscous fluid film of height $\hat h(\hat x,\hat t)$. The sheet is destabilized and attracted to the substrate by a van der Waals adhesion potential (Fig. 1a). Mechanical equilibrium requires that it the shape of the sheet is determined by the equation for the pressure\cite{Landau1986} $\hat p(\hat x,\hat t)$
\begin{equation}
\hat p(\hat x,\hat t)=B\hat h_{\hat x\hat x\hat x\hat x}(\hat x,\hat t)-\frac{A}{3\hat h^3(\hat x,\hat t)},
\label{eq:tfp}
\end{equation}
where $B\hat h_{\hat x\hat x\hat x\hat x}(\hat x,\hat t)$ is the elastic bending pressure due to long wavelength  deformations of the sheet with bending stiffness $B$ [N$\cdot$m], and $\frac{A}{3\hat h^3(\hat x,\hat t)}$ is the van der Waal adhesion pressure \cite{Israelachvili2011} with $A$ [N$\cdot$m] the Hamaker constant. Here, and elsewhere, the subscript denotes derivative i.e. $(\cdot)_{\hat x}=\partial (\cdot)/\partial {\hat x}$. Furthermore, we have assumed that the fluid film is very thin, so that its thickness $\hat h(\hat x,\hat t)/L\ll1$ and that it has a small slope $\hat h_{\hat x}(\hat x, \hat t)\ll1$, and finally that there are no inertial effects associated with the motion of the elastic sheet.

As the sheet approaches the substrate the intervening fluid is squeezed out, leading to non-uniform deformations of the sheet. In the small gap approximation, viscous forces dominate over inertial effects, so that we may use to the lubrication approximation for the description of the fluid motion \cite{Batchelor} that yields an evolution equation for the film height $\hat h(\hat x, \hat t)$ \cite{Batchelor} given by
\begin{equation}
\hat h_{\hat t}(\hat x,\hat t)=\left(\frac{\hat h^3(\hat x,\hat t)}{12\mu}\hat p_{\hat x}(\hat x,\hat t)\right)_{\hat x}=\left(\frac{\hat h^3(\hat x,\hat t)}{12\mu}\left(B\hat h_{\hat x\hat x\hat x\hat x}(\hat x,\hat t)- \frac{A}{3\hat h^3(\hat x,\hat t)}\right)_{\hat x}\right)_{\hat x}.
\label{eq:tfAdh}
\end{equation} 
$\mu$ is the fluid viscosity and we have assumed that there is no-slip of the fluid at both surfaces. Using (\ref{eq:tfp}), we find that the resulting sixth order nonlinear partial differential equation (\ref{eq:tfAdh}) couples the dynamics of the elastic sheet, the viscous forces in the lubricating film and the destabilizing intermolecular adhesion pressure.  

We scale the variables appearing in (\ref{eq:tfAdh}) as; $\hat x=\ell x=x \hat h_0(B/A)^{\frac{1}{4}}$, $\hat h(x,t)=\hat h_0 h(x,t)$ and $\hat t=\tau t=\frac{12 \hat h_0 \ell^2\mu}{A}t$, where $\tau=\frac{12 \hat h_0 \ell^2\mu}{A}$ is a viscous time scale, and introducing these scaled variables into (\ref{eq:tfAdh}) gives us the parameter free dimensionless elastohydrodynamic lubrication equation,
\begin{equation}
h_t(x,t)=\left({h^3(x,t)}h_{xxxxx}(x,t)- \frac{h_x(x,t)}{h(x,t)}\right)_x.
\label{eq:NDtf}
\end{equation} 

To complete the formulation of this initial value problem, we complement (\ref{eq:NDtf}) by an initial condition $h_0=h(x,t=0)=1-0.05\times\left(1+\cos(x)\right)$ corresponding to a film with a constant height that has a small blip at $x=0$, and six boundary conditions at the ends. Since the dynamics of interest in the touchdown region is far from the boundary, we use symmetry conditions at the two ends of the sheet, resulting in vanishing derivatives \cite{zhang:1999} i.e. $h_x=h_{xx}=h_{xxx}=0$. Note that these boundary conditions does not affect the dynamics near the touchdown at $x=x_C$, $t \rightarrow t_C$. 

Before solving (\ref{eq:NDtf}) numerically it is of interest to briefly consider the dispersion relation of the linearized version of (\ref{eq:NDtf}). Substituting $h(x,t)=1+\epsilon \exp(\lambda t+ikx)$ with $\epsilon \ll 1$ in (\ref{eq:NDtf}) with $k$ the wave number, $\lambda$ the growth rate, we find that $\lambda=k^2(1-k^4)$. Thus, we see that the elastic thin film is only unstable for long wavelength perturbations $k <1$, similar to what is known in capillary dominated flows. This suggests that when we solve (\ref{eq:NDtf}) numerically a minimal domain size $x\geq 2\pi/k_c$ is required, with $k_c=1$ the critical wave number.  

We solve (\ref{eq:NDtf}) using a second order finite difference discretization in space and a Gear method\cite{Gear} for the time stepping. In Fig. 1a we plot the evolution of the shape of the sheet and see that the solution appears to adopt a self-similar shape in the neighborhood of the touchdown point $x=x_C=0, t = t_C$, where the contact time $t_C$ is defined as the last time point before $h\leq 0$, just before the numerics breaks down.  
\begin{figure}
\centering
\begin{overpic}[width=0.9\linewidth]{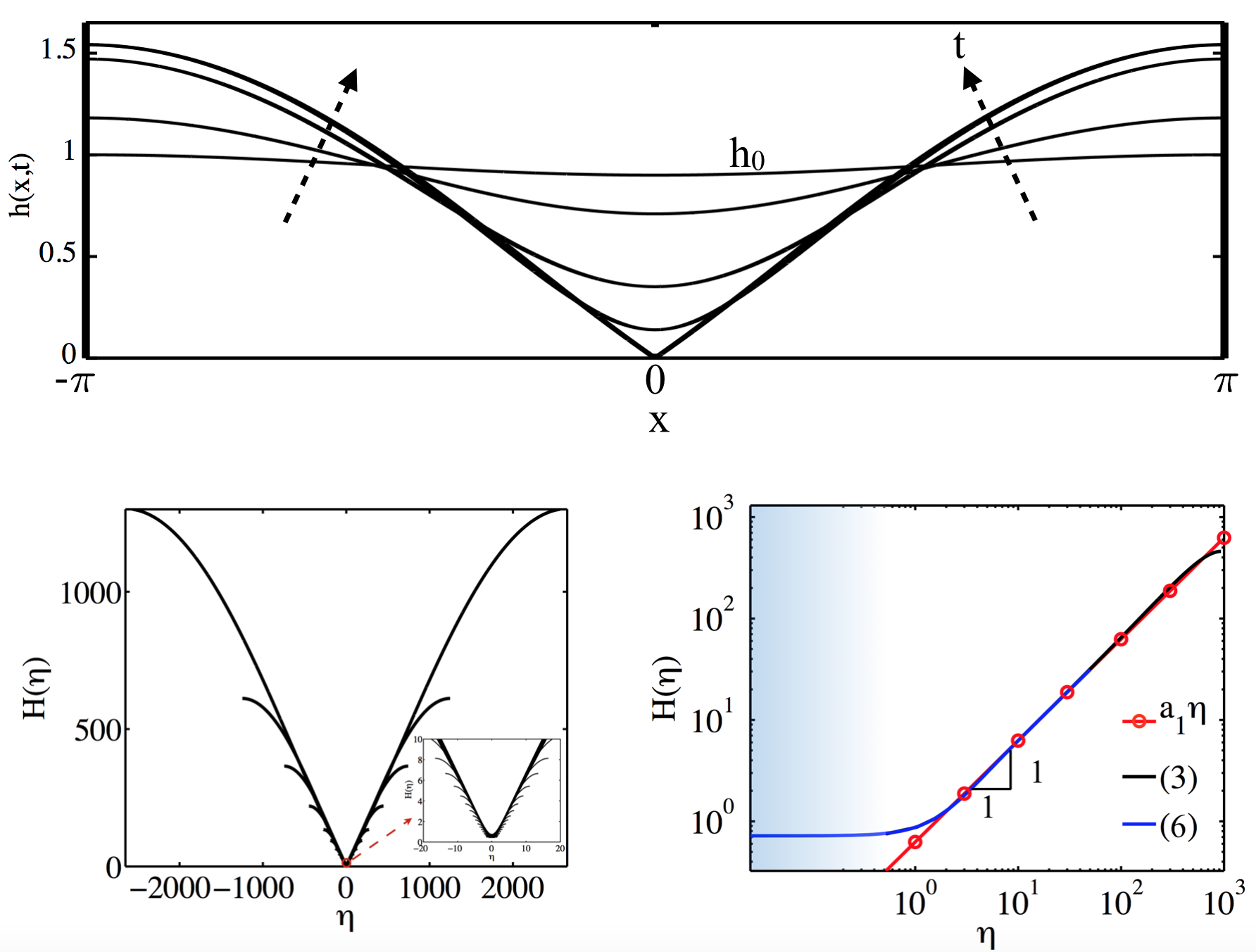}
\put(238,175){(a)}
\put(121,-7){(b)}
\put(356,-7){(c)}
\end{overpic}
\caption{(a) Numerical simulations of (\ref{eq:NDtf}) for the dynamics of touchdown showing the evolution of the sheet at five different time points $t=[0,~6.50,~7.98,~8.08,~8.09]$. The sheet is initialized with $h_0=h(x,0)=1-0.05\times\left(1+\cos(x)\right)$, and is seen to adopt a linear profile as $t\rightarrow t_C$ for $x\in[-2.5~2.5]$. Our simulations allow us to measure the contact point $x_C=0$ and contact time $t_C=8.0899$. (b) The shapes obtained by solving (\ref{eq:NDtf}) may be collapsed onto a universal self-similar shape by using the rescaled variables $H(\eta)=h(x,t)/(t_C-t)^{\frac{1}{3}}$,$\eta=(x-x_C)/(t_C-t)^{\frac{1}{3}}$, over more than two orders of magnitude in $H(\eta)$ and $\eta$. The inset in the lower right shows the self-similar collapse for $\eta\in [-20~20]$. (c) The rescaled height $H(\eta)$ obtained by solving the PDE (\ref{eq:NDtf}) and the similarity ODE (\ref{eq:sim}-7) (with $m=1$) shows that the height profiles adopt a shape $H(\eta)= a_1\eta$ as $\mid\eta\mid\rightarrow \infty$ ({\color{red}$\circ$}). The shaded area $\eta\leq 1$ illustrates the inner region near the contact point where the shape of the sheet arises from the balance between viscous forces, van der Waals adhesion pressure and elastic bending resistance. 
\label{fig:fig1}}
\end{figure}

Inspired by the self-similar shape of the sheet as $h\rightarrow 0$ in the neighborhood of $x=x_C, t=t_C$, we look for a self-similar solution of (\ref{eq:NDtf})  to understand the scaling behavior of the sheet in the immediate vicinity of touchdown. Adopting the similarity ansatz
\begin{equation}
h(x,t)=(t_C-t)^{\alpha}H(\eta),~~~\eta=\frac{x-x_C}{(t_C-t)^{\beta}},
\label{eq:ansatz}
\end{equation}
with $H(\eta)$ the scaled height, $\eta$ the similarity variable, and inserting (\ref{eq:ansatz}) into (\ref{eq:NDtf}) gives
\begin{equation}
(t-t_C)^{\alpha-1}(\alpha H(\eta)-\beta\eta H'(\eta))=\left((t-t_C)^{4\alpha-6\beta}H(\eta)^3H'''''(\eta) -(t-t_C)^{-2\beta}\frac{H'(\eta)}{H(\eta)}\right)'
\label{eq:powers}
\end{equation}
with $H'(\eta)\equiv \partial H(\eta)/\partial\eta$. Equating the exponents appearing on both sides of the equation, we find that  $\alpha-1=4\alpha-6\beta, \alpha-1=-2\beta$ so that $\alpha=\beta=1/3$ and are the same in axial symmetry\cite{axi}. Inserting this result in (\ref{eq:powers}) we find that the similarity variable $H(\eta)$ satisfies the ordinary differential equation
\begin{equation}
\frac{1}{3}\left(H(\eta)-\eta H'(\eta) \right)=\left(H^3(\eta)H'''''(\eta)- \frac{H'(\eta)}{H(\eta)} \right)'.
\label{eq:sim}
\end{equation}

To test the postulated self-similar scaling ansatz (\ref{eq:ansatz}) with $\alpha=\beta=1/3$, we re-plot the results obtained by solving the partial differential equation (\ref{eq:NDtf}) in Fig. 1a by rescaling $h(x,t)$ using the ansatz above. In  Fig. 1b, we see that the solution of (\ref{eq:NDtf}) converges rapidly to a self-similar form $H(\eta)$ over many orders of magnitude in $\eta$, confirming the validity of our similarity hypothesis.

The collapse of the time-dependent numerical results from (\ref{eq:NDtf}) by using the similarity transform in (\ref{eq:ansatz}) shows the self-similar shape of the sheet. Since this solution is also described by (\ref{eq:sim}), a natural question is if we can solve this equation directly and determine the self-similar shape associated with the touchdown event. This requires that we supplement (\ref{eq:sim}) with a set of boundary conditions. Noting that the numerical solution of (\ref{eq:NDtf}) in Fig. 1a is symmetric around the contact point $\eta=0$ suggests that odd derivatives of the solution vanish, i.e. $H'(0)=H'''(0)=H'''''(0)=0$. Furthermore, since the sheet height changes rapidly near the contact point i.e. $x=x_C$, $t\rightarrow t_C$, the far-field is quasi-static by comparison i.e. $h_t \approx 0$\cite{zhang:1999,Witelski:1999}. Thus, we expect that the far-field solution is independent of the evolution of the singularity as $t\rightarrow t_C$. This translates into a far-field Robin boundary condition that in similarity variables reads as 
\begin{equation}
\frac{1}{3}\left(H(\eta)-\eta H'(\eta)\right)= 0,~~~ \mid{\eta}\mid \rightarrow \infty. 
\end{equation}
We immediately see that the far-field solution is satisfied by $H(\eta)= a \eta +C$ as $\mid\eta\mid\rightarrow \infty$, and the integration constant $C=0$ for consistency. Plotting the results of the numerical solution of  (\ref{eq:NDtf}) shown in Fig. 1a in terms of the rescaled height $H(\eta)$ as $t\rightarrow t_C$ in logarithmic coordinates (Fig. 1c) confirms that $H(\eta)\approx \eta$ over a wide range of $\eta$, so that the constant $a$ is then interpreted as an asymptotic matching condition for the far-field associated with (\ref{eq:sim}), requiring us to prescribe an additional boundary condition. Since $H(\eta)\sim a \eta$ as $\eta \rightarrow \infty$, this leads to the following boundary conditions at infinity; $H'(\infty)=a$, $ H''(\infty)=H'''(\infty)=H''''(\infty)=0$, which together with the symmetry conditions at $\eta=0$ complete the formulation of the boundary value problem for $H(\eta)$ that satisfies (6).

\begin{table}
\centering
    \begin{tabular}{l l l l l l l}
        \hline
        \textbf{m} & \textbf{$a_m$} & \textbf{$H_m(0)$} & \textbf{$H''_m(0)$}&\textbf{$H''''_m(0)$} \\ \hline \hline
        PDE (3) & ~~~- & 0.708 & 0.324 & -0.183\\
        1 & 0.625 & 0.719 & 0.321 & -0.186\\
        2 & 0.290 & 0.452 & 0.031 & 0.190\\ 
        3 & 0.193 & 0.338 & 0.051 & -0.099\\ 
        4 & 0.147 & 0.283 & 0.0215 & 0.068\\ 
        5 & 0.119 & 0.245 & 0.0223 & -0.034\\\hline \hline
         \end{tabular}
\caption{Discrete solutions of the self-similar ODE (6-7) showing the values $a_m,H_m(0),H_m''(0),H_m''''(0)$. Comparison of these values with those obtained from the solution of the PDE (\ref{eq:NDtf}) show that it is consistent only with the fundamental self-similar solution $H_1(\eta)$.\label{tab:nd}}              
\end{table}

We solve the self-similar ODE (6-7) with the finite-difference-based boundary value solver BVP4C in Matlab\cite{MATLAB} on a finite size domain $\eta=50$, with the boundary conditions; $H'(0)=H'''(0)=H'''''(0)=H''(50)=H'''(50)=H''''(50)=0$, $H'(50)=a$, noting that our results are insensitive to changes in the domain size. We find that we get a discrete family of solutions for (\ref{eq:sim}-7) similar to those seen in the problem of capillary rupture \cite{Bernoff:1998}. The subscript $m$ is used to distinguish the different discrete solutions of $H(\eta)$and $a$ with
\begin{equation}
H(\eta)\equiv H_m(\eta),~~~~~~a\equiv a_m.
\end{equation} 
The numerical procedure starts with an initial guess for the height at contact $H_0(0)$, together with a guess of the initial solution $H_0(\eta)=H_0(0)+a_0 \eta$. We systematically vary the initial guesses $H_0(0)$ and $a_0$, allowing us to identify the first thirteen solutions $m=1, 2, ..., 13$ for $H_m(\eta)$ satisfying (\ref{eq:sim}-7), see Fig. 2a .

Comparison of these solutions of ODE (\ref{eq:sim}-7) with the solution of  the time-dependent partial differential equation (\ref{eq:NDtf}) shows that only the fundamental solution associated with $m=1$ agrees with the PDE in (\ref{eq:NDtf}), as shown in Fig. 2a. 

To understand why only the fundamental solution $m=1$ is consistent with the time-dependent solution of (\ref{eq:NDtf}) we qualitatively evaluate the stability of these self-similar solutions, using $H_m(\eta)$ as the initial condition for the time dependent PDE (\ref{eq:NDtf}). In Fig. 2b, we show the result of our simulation of (\ref{eq:NDtf}) with $h(x,0)=H_m(\eta)$, $m=3$. The time and position of contact shifts and similar results arise for all values of $m>1$. Similar observations have been made for capillary thin film rupture \cite{Bernoff:1998}. Rescaling the results with (4) shows that the solution converges to that associated with the shape of the fundamental solutions $m=1$ in the vicinity of touchdown (Fig. 2c). While a detailed stability analysis of the problem is necessary to classify the different unstable modes, we postpone this to the future and instead give a simple heuristic explanation next. 

Starting with the different discrete solutions of the similarity equation $H_m(\eta)$, we see that for $m>1$ there are small pressure oscillations that are defined by the bending pressure $H_m''''(\eta)$ in the neighborhood of $\eta=0$. As $m$ increases $a_m$ decreases and there is a subsequent flattening of the film (Table 1, Fig. 2a). The number of maxima/minima increase linearly with $m$ as can be seen in Fig. 2d. Since these correspond to short wavelength modes of the sheet, they are damped out much more strongly than the longest wavelength mode corresponding to $H_1(\eta)$ which is balanced by the van der Waals adhesion pressure $H^{-3}(\eta)$ and the viscous stress, and this self-similar solution is the only dynamically stable one that survives.

\begin{figure}[h]
\centering
\begin{overpic}[width=0.9\linewidth]{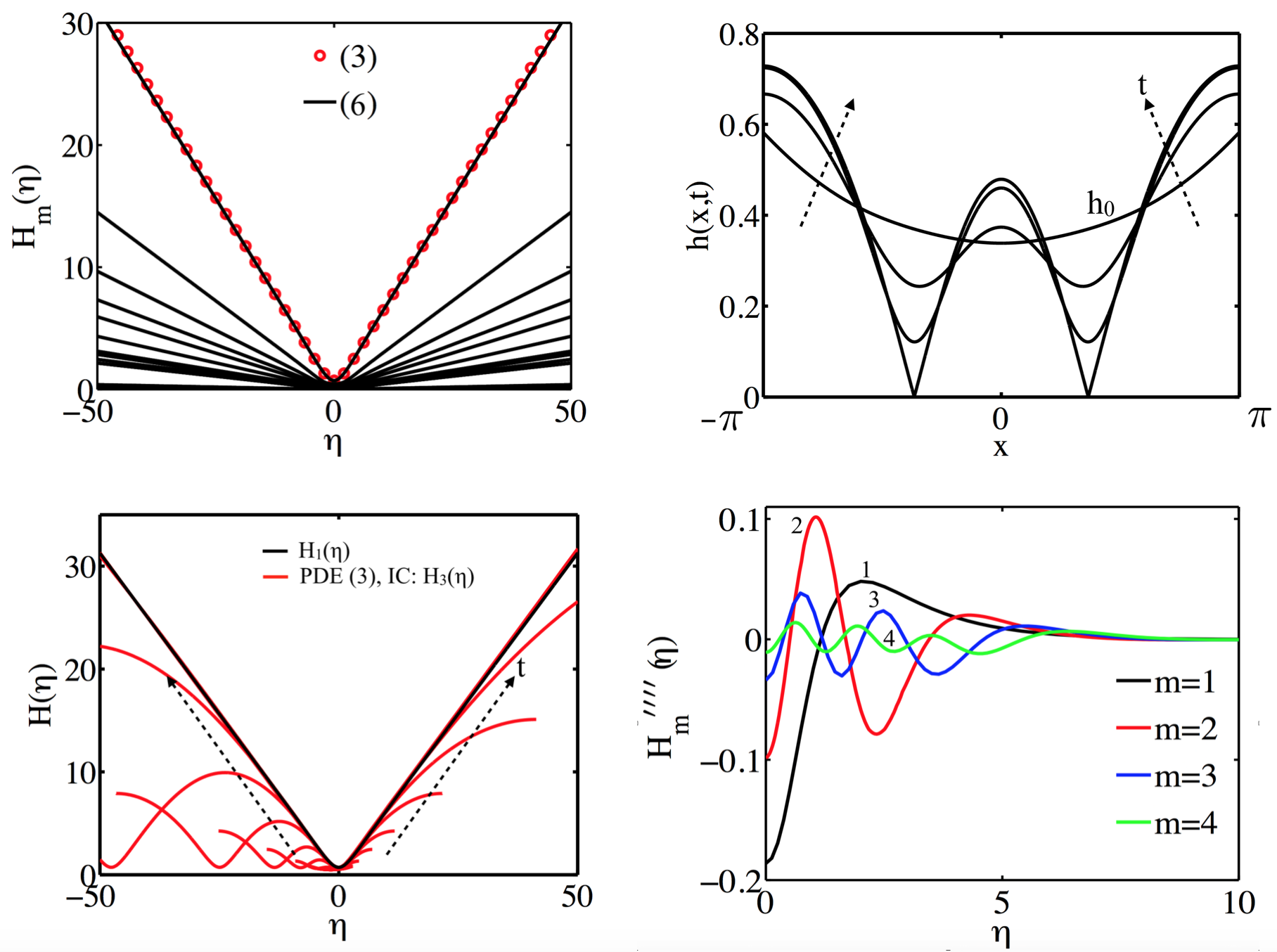}
\put(117,168){(a)}
\put(358,168){(b)}
\put(119,-9){(c)}
\put(358,-9){(d)}
\end{overpic}
\caption{(a) Comparison between the self-similar shape of the sheet predicted by the time-dependent PDE (\ref{eq:NDtf}) and the first thirteen solutions $H_m(\eta); m\in [1-13]$ of the similarity ODE (\ref{eq:sim}-7). We note that only the first fundamental solution $H_1(\eta)$ of (6-7) is consistent with that of the PDE (\ref{eq:NDtf}) ({\color{red}$\circ$}). (b) Numerical simulations of the PDE (\ref{eq:NDtf}) using the initial condition associated with the self-similar shape $h_0=h(x,t=0)=H_3(\eta)$ shows the evolution of the shape of the sheet for four different time points $t=[0,~0.28,~0.336,~0.341]$. We find that the numerically measured contact time is $t_C=0.3412$ and the contact points are $x_C=[-1.48,1.48]$. (c)  By using the rescaled variables $H(\eta)=h(x,t)/(t_C-t)^{\frac{1}{3}}$,$\eta=(x-x_C)/(t_C-t)^{\frac{1}{3}}$ with $x_C=1.48$ corresponding to the contact on the right side, we see that the shapes of the sheet obtained in Fig. 2a collapse onto the universal shape of the fundamental solution of (6-7) $H_1(\eta)$ (shown in red). The deviations away from the contact point are due to the effects of the second touchdown at $x_c=-1.48$. (d) To distinguish the different discrete solutions $H_m(\eta)$ satisfying (6-7) we plot the scaled bending pressure $H_m''''(\eta)$, and see that the number of maxima/minima scales with $m$. Thus, for $m>1$, the wavelength of the pressure oscillations decreases as $m$ increases; these solutions are therefore strongly damped out in comparison with the fundamental solution of $H_1(\eta)$ (6-7), which alone survives as we approach touchdown.
\label{fig:groupedfig2}}
\end{figure}
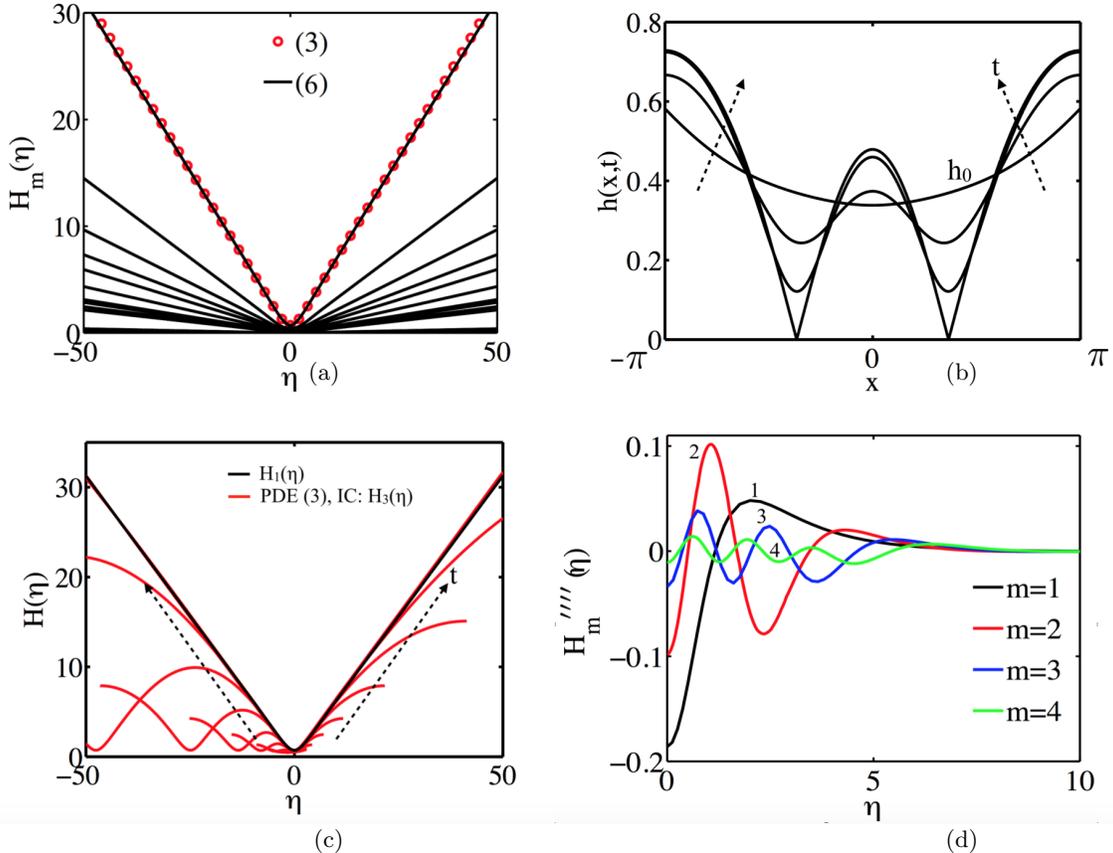

We conclude with a few remarks on the approximations inherent in our approach.  As the minimal height becomes comparable to the molecular dimension $\approx 2.5$\AA, the continuum description that we have used breaks down. Our simulations show that the self-similar shape is adopted already at $h(x,t)\approx0.3$ (Fig. 1b) and choosing a typical value for the initial height $\hat h_0=300$nm gives a height $\approx 100$nm which is far from this molecular height and within experimental realm using total internal reflection microscopy imaging. During touchdown, the minimal height $h_{min}(t)$ and the horizontal deformation length $\ell(t)=h_{min}(t) (B/A)^{\frac{1}{4}}$ becomes dynamic. If $\ell(t)$ is comparable to the thickness of the sheet $b\sim h_{min}(t) (B/A)^{\frac{1}{4}}$, three-dimensional elastic effects may become relevant. However, since our small slopes approximation holds throughout time as $h_x\sim h_{min}(t)/\ell(t)\sim (A/B)^{\frac{1}{4}}\ll1$ with $B\gg A$ e.g. $A=10^{-20}$N$\cdot$m, $B=1$N$\cdot$m the plate theory that we have used here remain reasonable. Finally, as the sheet and fluid velocity diverges near the singularity, this can lead to inertial effects, particularly as the reduced Reynolds number $Re\equiv\frac{\rho U h_0^2}{L\mu}\sim (t_C-t)^{-\frac{1}{3}}$ is also singular as $t\rightarrow t_C$. By assuming $Re\approx O(1)$, we get a scaling estimate for the height at which inertia becomes significant $\hat h_I\sim \frac{\rho A^{\frac{3}{2}}}{12B^{\frac{1}{2}}\mu^2}$ e.g. $\mu=10^{-3}$ Pa$\cdot$s, $\rho=10^3$ kg/m$^3$, giving $\hat h_I<2.5$\AA~and the approximation of neglecting inertial effects remains valid. 

Our analysis reveal a new class of singular flows linking elasticity, hydrodynamics and adhesion, relevant to contact between a thin elastic sheet and an adherent surface. A combination of numerical simulations and similarity analysis shows that the dynamic height and horizontal deformation of the sheet in the process of contact show a power-law form. Analysis of the governing equations in similarity coordinates further allows us to determine the universal self-similar shape of the elastic sheet in the neighborhood of touchdown. A qualitative analysis of the discrete set of self-similar film shapes that satisfy the similarity ODE shows that  the solution with the least oscillatory bending pressure is the one that is consistent with the time-dependent numerical simulation, while others with short-wavelength pressure oscillations are strongly damped. A natural next step is to experimentally test the regime of applicability of our results. 

\bibliographystyle{aipsamp}

\begin{thebibliography}{100}
\bibitem{Batchelor}
Batchelor,  G. K. {\em An Introduction to Fluid Dynamics (Cambridge University Press)} (1967).

\bibitem{Barenblatt}
\newblock Barenblatt, G. I. {\em Similarity, Self-Similarity, and Intermediate Asymptotics (Consultants Bureau)}, 1979.

\bibitem{eggers:1999}
\newblock Eggers J. ''Nonlinear dynamics and breakup of free-surface flows`` {\em Reviews of Modern Physics} {\bf 69}, 865--929 (1997).

\bibitem{eggers:2009}
\newblock Egger, J. and Fontelos M. A. ''The role of self-similarity in singularities of partial differential equations`` {\em Nonlinearity} {\bf 22}, R1--44 (2009).

%
\bibitem{jain:1976}
\newblock Jain, R. K. and Ruckstein E. ''Stability of stagnant viscous films on a solid surface`` {\em Journal Colloid Interface Science} {\bf 54}, 108--116 (1976).

\bibitem{william:1982}
\newblock Williams M. B. and Davis S. H. '' Non-linear theory of film rupture`` {\em Journal of Colloidal and Interface Science} {\bf 99}, 220--228 (1982).
%
\bibitem{Bruelbacht:1988}
\newblock Bruelbacht, J. P., Bankoff, S. G. and Davis, S. H. ''Nonlinear stability of evaporating/condensing liquid films`` {\em Journal of Fluid Mechanics} {\bf 195}, 463--494 (1988).
%

\bibitem{Bernoff:1998}
\newblock Bernoff A. J., Bertozzi, A. L. and Witelski T. P. ''Axisymmetric surface diffusion: dynamics and stability of self-similar pinchoff`` {\em Journal of Statistical Physics} {\bf 93}, 725--776 (1998).

\bibitem{zhang:1999}
\newblock Zhang W. and Lister J. R. ''Similarity solutions for van der Waals rupture of a thin film on a solid substrate`` {\em Physics of Fluids} {\bf 11}, 2454---2462 (1999).

\bibitem{Witelski:1999}
\newblock Witelski T. P. and Bernoff A. J. ''Stability of Self-similar Solutions for Van der Waals Driven Thin Film Rupture`` {\em Physics of Fluids} {\bf 11}, 2443--2445 (1999).

\bibitem{Witelski:2000}
\newblock Witelski T. P. and Bernoff A. J. ''Dynamics of three-dimensional thin film rupture`` {\em Physica D} {\bf 147}, 155--176 (2000).

\bibitem{Vaynblat:2001}
\newblock Vaynblat D., Lister, J. R. and Witelski, T. P. ''Rupture of thin viscous films by van der Waals forces: Evolution and self-similarity`` {\em Physics of Fluids} {\bf 13}, 1130---1140 (2001).

%


\bibitem{Witelski:2010}
\newblock Bernoff A. J. and Witelski T. P. ''Stability and dynamics of self-similarity in evolution equations`` {\em Journal of Engineering Mathematics} {\bf 66}, 11--31 (2010).

%
%

%
%

%




\bibitem{craster:2009}
\newblock Craster R. V. and Matar O. K. ''Dynamics and stability of thin liquid films`` {\em Reviews of Modern Physics} {\bf 81}, 1131--1198 (2009).




\bibitem{Landau1986}
\newblock Landau, L. D. and Lifshitz, E. M. {\em Theory of Elasticity (Elsevier)} (1986).

\bibitem{Israelachvili2011}
\newblock Israelachvili, J. N. {\em Intermolecular and Surface Forces 3rd Edition (Elsevier)} (2011).

%
 


\bibitem{axi}
\newblock The axisymmetric elastohydrodynamic thin film equation is given by $r h_t=\left(h^3(rh_{rrrr}+2h_{rrr}-h_{rr}/r + 1/r^2 h_r)_r+ h^3(r/h^3)_r\right)_r$ and introducing (\ref{eq:ansatz}) also yields $\alpha=\beta=1/3$ for the two-dimensional case. However, it is likely that there will be non-axisymmetric instabilities during touchdown in this case. 
\bibitem{Gear}
Gear, C. W. {\em IEEE Transactions on Circuit Theory} {\bf 18}, 89-95 (1971).     

\bibitem{MATLAB}
\newblock MATLAB version 8.2.0.701 (R2013b). {\em The MathWorks Inc.} {\bf 49}, (2013).



\end{thebibliography}

\end{document}